\documentclass[a4paper,twoside]{article}

\usepackage{epsfig}
\usepackage{subcaption}
\usepackage{calc}
\usepackage{amssymb}
\usepackage{amstext}
\usepackage{amsmath}
\usepackage{amsthm}
\usepackage{multicol}
\usepackage{pslatex}
\usepackage{apalike}
\usepackage{SCITEPRESS}     

\begin{document}

\title{Metadata Management for Textual Documents in Data Lakes}

\author{\authorname{Pegdwend\'e N. Sawadogo\sup{1}, Tokio Kibata\sup{2} and J\'er{\^o}me Darmont\sup{1}}
\affiliation{\sup{1}Universit\'e de Lyon, Lyon 2, ERIC EA 3083 \\ 
5 avenue Pierre Mend\`es France, F69676, Bron, France}
\affiliation{\sup{2}Universit\'e de Lyon, Ecole Centrale de Lyon \\ 
36 avenue Guy de Collongue, F69134, Ecully, France}
\email{pegdwende.sawadogo@univ-lyon2.fr, tokio.kibata@ecl16.ec-lyon.fr, jerome.darmont@univ-lyon2.fr}
}

\keywords{Data lakes, Textual documents, Metadata management, Data ponds}

\abstract{Data lakes have emerged as an alternative to data warehouses for the storage, exploration and analysis of big data. In a data lake, data are stored in a raw state and bear no explicit schema. Thence, an efficient metadata system is essential to avoid the data lake turning to a so-called data swamp. Existing works about managing data lake metadata mostly focus on structured and semi-structured data, with little research on unstructured data. Thus, we propose in this paper a methodological approach to build and manage a metadata system that is specific to textual documents in data lakes. First, we make an inventory of usual and meaningful metadata to extract. Then, we apply some specific techniques  from the text mining and information retrieval domains to extract, store and reuse these metadata within the COREL research project, in order to validate our proposals.}

\onecolumn \maketitle \normalsize \vfill

\section{\uppercase{Introduction}}
\label{sec:Introduction}
The tremendous growth of social networks and the Internet of Things (IoT) brings various organisms exploit more and more data. Such amounts of so-called big data are mainly characterized by high volume, velocity and variety, as well as a lack of veracity, which together exceed the capacity of traditional processing systems \cite{Miloslavskaya2016}. 
To tackle these issues, Dixon introduced the concept of data lake, a large repository of raw and heterogeneous data, fed by external sources and allowing users to explore, sample and analyze the data \cite{Dixon2010}. 

In a data lake, original data are stored in a raw state, without any explicit schema, until they are queried. This is known as schema-on-read or late binding 
\cite{Fang2015,Miloslavskaya2016}. However, with big data volume and velocity coming into play, the absence of an explicit schema can quickly turn a data lake into an inoperable data swamp \cite{Suriarachchi2016}. Therefore, metadata management is a crucial component in data lakes \cite{Quix2016}. An efficient metadata system is indeed essential to ensure that data can be explored, queried and analyzed.   

Many research works address metadata management in data lakes. Yet, most of them focus on structured and semi-structured data only \cite{Farid2016,Farrugia2016,Madera2016,Quix2016,Klettke2017}. Very few target unstructured data, while the majority of big data \textit{is} unstructured and mostly composed of textual documents \cite{Miloslavskaya2016}. Thus, we propose a metadata management system for textual data in data lakes. 

Our approach exploits a subdivision of the data lake into so-called data ponds \cite{Inmon2016}. Each data pond is dedicated to a specific type of data (i.e., structured data, semi-structured data, images, textual data, etc.) and involves some specific data preprocessing. Thus, we propose in this paper a textual data pond architecture with processes adapted to textual metadata management. We notably exploit text mining and information retrieval techniques to extract, store and reuse metadata. 

Our system allows two main types of analyses. 
First, it allows OLAP-like analyses, i.e., documents can be filtered and aggregated with respect to one or more keywords, or by document categories such as document MIME type, language or business category. Filter keys are comparable to a datamart's dimensions, and measures can be represented by statistics or graphs. 
Second,  similarity measures between documents can be used to automatically find clusters of documents, i.e., documents using approximately the same lexical field, or to calculate a document's centrality. 
We demonstrate these features in the context of  the COREL research project.  

Our contribution is threefold.
1) We propose the first thorough methodological approach for managing unstructured, and more specifically textual, data in a data lake.
2) We introduce a new type of metadata, global metadata, which had not been identified as such in the literature up to now.    
3) Although we articulate existing techniques (notably standards) to build up our metadata management system, adaptations are required. We especially combine a graph model and a data vault \cite{Linstedt2011} for metadata representation, and extend an XML representation format for metadata storage.

The remainder of this paper is organized as follows. Section~\ref{sec:RelatedWorks} surveys the research related to metadata management in data lakes, and especially textual metadata issues. Section~\ref{sec:MetadataManagement} presents our metadata management system. In Section~\ref{sec:POC}, we apply our approach on the COREL 
data lake as a proof of concept. Finally,  Section~\ref{sec:Conclusion} concludes the paper and gives an outlook on our research perspectives.

\section{\uppercase{Related Works}}
\label{sec:RelatedWorks}

\subsection{Metadata Management in Data Lakes}

\subsubsection{Metadata Systems}
\label{sec:mdsys}
There are two main data lake architectures, each adopting a particular approach to organize the metadata system.  
We call the first architecture storage-metadata-analysis \cite{Stein2014,Quix2016,Hai2016}, where the metadata system is seen as a global component for the whole dataset. Every analysis or query is then performed through this component.

The second architecture structures the data lake into data ponds. A data pond is a subdivision of a data lake, dealing with a specific type of data \cite{Inmon2016}. In this approach, storage, metadata management and querying are specific to each data type (voluminous, structured data from applications; velocious, semi-structured data from the IoT; various, unstructured textual documents). This organization helps conform to data specificity. 

\subsubsection{Metadata Generation}
\label{sec:mdgen}

There are many techniques in the literature to extract metadata from a data lake. For instance, generating data schemas or column names make formulating queries and analyses easier \cite{Quix2016,Hai2016}. In the same line, integrity constraints can be deduced from data \cite{Farid2016,Klettke2017}. However, such operations are not applicable to textual data. 

There are three main ideas in the literature for building and managing metadata that are appropriate for textual data. 
The first proposal consists in indexing the documents. 
This is notably applied in the CoreDB data lake to support a keyword querying service \cite{Beheshti2017}. 

The second idea, called semantic annotation \cite{Quix2016}, semantic enrichment \cite{Hai2016} or semantic profiling \cite{Ansari2018}, adds a context layer to the data, which defines the meaning of data. This is done using World Wide Web Consortium standards such as OWL \cite{Laskowski2016}. It is used for enriching a couple of data lakes \cite{Terrizzano2015,Quix2016}. 

Finally, the third proposal applies a textual disambiguation process before document ingestion in the data lake \cite{Inmon2016}. Textual disambiguation consists in, on one hand, providing context to the text, e.g., using taxonomies; and on the other hand, transforming the text into a structured document.  

\subsection{Discussion}

Current metadata management techniques for textual data (Section~\ref{sec:mdgen}) are all relevant. Each brings in a crucial feature: indexing permits filtering data with one or more keywords; semantic enrichment complements data with domain-specific information; and textual disambiguation makes the automatic processing of textual data easier. 

To achieve an efficient metadata system, an idea can be to take advantage of these three techniques, combining them into a global metadata management system. However, it is not so simple. 

A first problem is that textual disambiguation implies that the original data are transformed before being fed to the data lake \cite{Inmon2016}. Thus, raw data are lost during this process, which contradicts the definition of data lakes \cite{Dixon2010}.

A second problem also concerns textual disambiguation. We need to define concrete examples of structured formats in which textual data can be converted to be easily analyzed. This remains an open issue as far as we know.

A third problem is that seeking for semantic enrichment via all possible semantic technologies seems illusory. There are indeed many semantic enrichment methods, which limits user flexibility.

Eventually, current metadata management techniques do not consider an important type of metadata, i.e., relational metadata. 
Relational metadata, also called inter-dataset metadata (Section~\ref{sec:InterMetadata}), express tangible or intangible links between datasets or data ponds \cite{Maccioni2017}. 
Obtaining relational metadata indeed allow advanced analyses such as centrality 
and community detection proposed in a semi-structured data context \cite{Farrugia2016}, and that could be extended to unstructured documents. In the context of textual documents, a community typically means sharing a  lexical field, and can serve to automatically classify documents by topics.

\section{\uppercase{Textual Metadata Management System}}
\label{sec:MetadataManagement}

\subsection{Metadata Identification}
\label{sec:mdid}

A categorization of metadata in data lakes \cite{Maccioni2017} distinguishes intra-dataset (Section~\ref{sec:IntraMetadata}) and inter-dataset metadata (Section~\ref{sec:InterMetadata}). In addition, we propose a third category:  global metadata (Section~\ref{sec:GlobalMetadata}).

\subsubsection{Intra-dataset Metadata}
\label{sec:IntraMetadata}

This category is composed of metadata concerning only one dataset (a textual document in our context) in a data pond. Three subcategories are relevant to textual documents. 
The first category is made of \textit{properties} that take the form of key-value pairs and provide information about  data  \cite{Quix2016}, e.g., document creation date, document creator, document length, etc.

The second category of intra-dataset metadata is called \textit{previsualization metadata}. They consist of a summary of the document through a visualization or a set of descriptive tags \cite{Halevy2016}. Their role is to provide users with an idea of the document's contents.

Finally, the last category concerns \textit{version metadata}. These metadata are inspired from 
the idea of  keeping different versions of each document in the data pond \cite{Halevy2016}, e.g., a lemmatized version, a version without stopwords, etc.

To enhance version metadata, we introduce the notion of \textit{presentation metadata}. 
Presentation metadata are obtained by applying two operations to the original data. The first is a transformation operation to get a cleaned or enriched version of a document, e.g., stopwords removal, lemmatization, etc. 

The second operation consists in giving a structured format to the transformed document to actually achieve the presentation as, e.g., a bag of words, a term-frequency vector or a TF-IDF vector. 

The generation of presentation metadata is similar to text disambiguation \cite{Inmon2016}, except that our proposal does \textit{not} involve any data loss. Original data are retained.

\subsubsection{Inter-dataset Metadata}
\label{sec:InterMetadata}

Inter-dataset or relational metadata specify relationships between different documents \cite{Maccioni2017}. There are two subcategories of inter-dataset metadata: physical and logical links~\cite{Farrugia2016}.

Physical links represent a clear (tangible) connection between documents. Such connections are typically induced by the belonging to some natural or business clusters, e.g., creation by the same person or belonging to the same business category, etc. 
In contrast, logical (intangible) links highlight similarities between documents from their intrinsic characteristics, such as common word rate or inherent topics.

\subsubsection{Global Metadata}
\label{sec:GlobalMetadata}
This category of metadata covers the whole data lake. Global metadata may be used and reused to enrich documents or to perform more advanced analyses. They generally include semantic data such as thesauri, dictionaries, ontologies, etc. 

Such metadata can be exploited to create enriched or contextualized documents through semantic enrichment \cite{Hai2016}, annotation~\cite{Quix2016} or profiling \cite{Ansari2018}. This process is then the first part of presentation metadata generation (Section~\ref{sec:IntraMetadata}) . 

Global metadata can also serve to enrich data querying, e.g., a thesaurus can be used to expand a keyword-based query with all the synonyms of the initial keywords. 

Eventually, semantic resources such as ontologies and taxonomies can also help classify documents into clusters \cite{Inmon2016,Quix2016}. For example, Inmon uses two taxonomies of positive and negative terms  to detect whether a text brings out a positive or negative sentiment, respectively.

\subsection{Metadata Representation}
A data lake or data pond can be viewed as a graph, where nodes represent documents and edges express connections or similarities between documents \cite{Farrugia2016,Halevy2016}. Such a representation allows to discover communities or to calculate the centrality of nodes and, thus, to distinguish  documents sharing the same lexical field from those with a more specific vocabulary 
 \cite{Farrugia2016}. We adopt this approach because it is relevant to inter-dataset metadata (Section~\ref{sec:InterMetadata}). 

Moreover, to handle the changing number and form of intra-dataset metadata, we combine the graph view of metadata with data vault modeling. 
Data vaults are alternative logical models to data warehouse star schemas that, unlike star schemas, allow easy schema evolution \cite{Linstedt2011}.  They have already been adopted to represent a data lake's metadata \cite{Nogueira2018}, but retaining a relational database structure that does not exhibit an explicit graph representation.

In contrast, we associate data vault satellites representing intra-dataset metadata with graph nodes.
In our context, a satellite stores descriptive information, i.e., a set of attributes associated with one specific document. 
Yet, a document may be described by several satellites \cite{Hultgren2016}.
Then, with the help of this one-to-many relationship, we can easily associate any new intra-dataset metadata with any document by creating new satellites attached to the document's node.

\begin{figure*}[hbt]
 \centering
 \includegraphics[width=9cm]{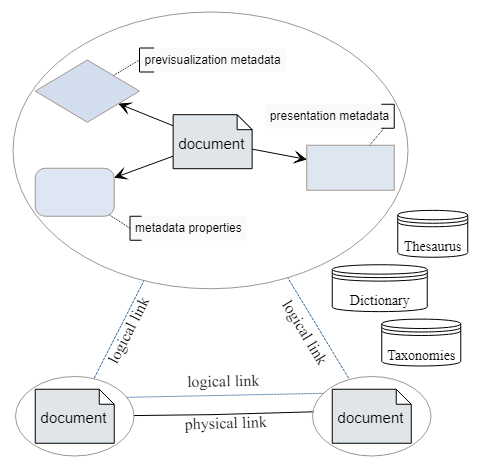} 
 \caption{Sample metadata representation}
 \label{fig:MetadataGraphPres}
\end{figure*}

Figure~\ref{fig:MetadataGraphPres} shows an example of metadata representation for three documents. Visualization metadata, presentation metadata and properties are associated with each document. The association arrow from a document to metadata indicates that it is possible to associate several instances of every type of metadata with a document. 

Inter-dataset metadata are subdivided into logical links represented by dotted edges and physical links represented by solid edges. It is also possible to define several instances of inter-dataset metadata. The only constraint that we introduce is that each logical link, when defined, should be generated between every couple of documents, which is not the case of physical links. 
That is, we assume that there is always some similarity 
between two documents. The question is then to know the strength of this similarity. 
For physical links, the question is simply whether the link exists.

Finally, global metadata are not as such part of the graph, because they are not directly connected to any document. This is why they are isolated.

\subsection{Metadata Storage}

To efficiently store the metadata identified in Section~\ref{sec:mdid}, we adopt the idea to associate an XML metadata document with each document. This XML document serves as the textual document's ``identity card" and permits to store and retrieve all its related metadata.
This approach has notably been used to build a data preservation system for the French National Library \cite{Fauduet2010}. 
Each digital document is ingested in their system as a set of information pieces represented by an XML manifest (Section~\ref{sec:manifest}) that can be viewed as a metadata package \cite{Fauduet2010}.

Depending on the type of metadata, we propose three storage modes: integrally within, partially within and independent 
from a manifest (Sections~\ref{sec:atomic}, \ref{sec:nonatomic} and \ref{sec:relational}, \textit{irrespectively}).

\subsubsection{XML Manifest Structure}
\label{sec:manifest}

The manifest XML document associated with each textual document is composed of three sections, each dedicated to a specific metadata type. The first two sections are defined by the Metadata Encoding \& Transmission Standard (METS) \cite{TLoC2017}.  

The first section, named \textit{dmdSec}, stores atomic metadata in a \textit{mdWrap} subsection (Section~\ref{sec:atomic}). 
The second section is another  \textit{dmdSec} section where non-atomic metadata are referenced by a pointer in \textit{mdRef} XML elements (Section~\ref{sec:nonatomic}). 

The third section is our proposal, which we name \textit{prmSec} for physical relational metadata section. It stores physical links in \textit{prm} elements.

\begin{figure*}[hbt]
 \centering
 \includegraphics[width=12cm]{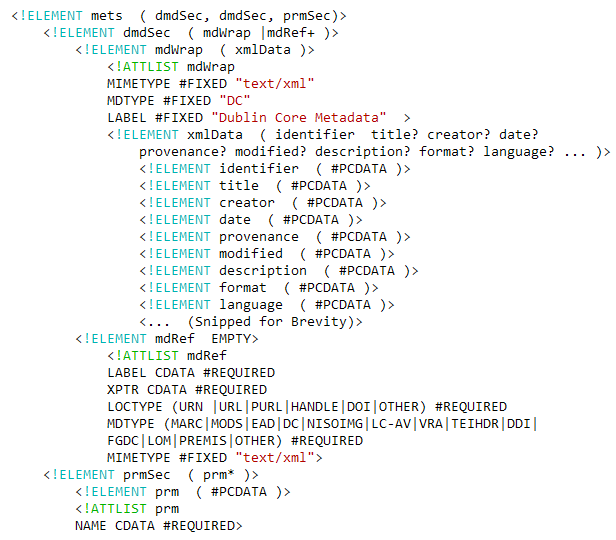} 
 \caption{Document manifest DTD}
 \label{fig:DatasetManifestDTD}
\end{figure*}

Figure~\ref{fig:DatasetManifestDTD} shows the document manifest schema as an XML DTD, which we choose because it is more humanly understandable than an XML Schema. 
Eventually, to make exploring and querying metadata easier, we strongly advocate for storing the set of XML manifest documents into an XML DBMS.

\subsubsection{Atomic Metadata Storage}
\label{sec:atomic}

Metadata in an atomic or near-atomic form are 
directly included in the XML manifest. It is generally the case of properties, e.g., the document's identifier, its creation or last modification timestamp, its creator's name, etc.
More precisely, such metadata are stored in a \textit{dmdSec} section/\textit{mdWrap} subsection. In this subsection, atomic metadata are represented by XML elements from the (standard) Dublin core namespace \cite{DublinCore}. 
However, the proposed namespace can easily be replaced by a customized namespace.

\subsubsection{Non-atomic Metadata Storage}  
\label{sec:nonatomic}

Several types of metadata bear a format that requires a specific storage technology. Therefore, such metadata cannot be stored internally in the manifest like atomic and near-atomic metadata. They are thus stored in a specific format in the filesystem. 
Yet, to make the retrieval of this kind of metadata easier, they are referenced in the second \textit{mdRef} section of the manifest, through URIs.   

Presentation and previsualization metadata fall in this category. The original data must also be referenced this way, to allow users easily retrieve the raw data whenever necessary.

\begin{figure}[hbt]
 \centering
 \includegraphics[width=6.75cm]{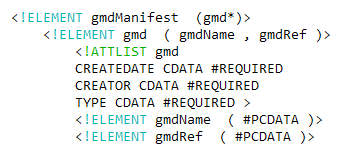} 
 \caption{Global manifest DTD}
 \label{fig:GlobalManifestSchema}
\end{figure}

Global metadata are also stored using specific technologies and externally referred to. However, as they do not concern a specific document, they cannot be included in a document manifest. Thus, we propose the use of a special XML manifest document for all global metadata. This manifest contains a set of global metadata for which the name, location and type (e.g., thesaurus, dictionary, stopwords, etc.) 
are specified. The global manifest's schema is provided in Figure~\ref{fig:GlobalManifestSchema} as a DTD.

\subsubsection{Relational Metadata}
\label{sec:relational}

Inter-dataset metadata have some specific requirements because they concern two documents. It is thus difficult to include them in a document manifest. 
Hence, we propose to include physical links in the \textit{prmSec} section of all concerned document manifests. 
The physical link's name must be specified. Its value represents a cluster name, e.g., for a physical link expressing the document's language, the link's name can be "language" and possible values "fr", "en", "de", etc.  
Thus, clusters can be reconstituted by grouping all documents of equal value with respect to a specific physical link.

Unlike physical links, logical links cannot be defined internally within a manifest, since they materialize between each couple of documents with a given strength.
Thus, to store logical links, we propose to use a graph DBMS such as Neo4j \cite{Neo4j}, which allows to easily store link strengths and to conform to the graph view from Figure~\ref{fig:MetadataGraphPres}.  
In this approach, each node represents a document by the same identifier as the one in the manifest. Each type of logical link is then defined between every couple of documents as an edge labeled with its name and strength.

\subsection{Metadata Extraction}

Several techniques can be applied to generate metadata. We categorize them with respect to the type of metadata in the following sections.
Moreover, here, we redefine intra and inter-dataset metadata as intra and inter-document metadata, respectively, to fit our textual document data pond context.

\subsubsection{Intra-document Metadata Extraction}
Properties can be obtained from the filesystem, e.g., document name, size, length, location or date of last modification \cite{Quix2016}. Once metadata are extracted, they are inserted into the manifest document, into the corresponding element from the Dublin Core's namespace.
Moreover, to conform to our metadata storage system, an ID must be generated for each document.  ID generation can be based either on the document's URI, if any \cite{Suriarachchi2016}, or on the document manifest's generation timestamp. 
We choose to use the manifest's generation timestamp to avoid possible conflicts resulting from moving or changing the name of the document.

Then, there are two ways to generate previsualization metadata. The first is manually adding a set of tags to a document, in the form of atomic metadata. The second is automatically generating metadata by applying, e.g., topic modeling techniques.

Presentation metadata generation requires two advanced operations. The first operation consists in either data cleaning (e.g., stopwords removal, lemmatization, filtering on a dictionary, etc.) or data enrichment (e.g., adding context using taxonomies, translations, etc.). 
At this stage, a transformed version of the document is obtained. The second operation, e.g., presentation as a bag of words or a term-frequency vector, is then applied on the transformed document to generate presentation metadata. 

Generated metadata can then be stored either in the filesystem, possibly with an indexing technology such as Elasticsearch \cite{Elasticsearch} on top, or within a DBMS. 
We opt for a hybrid storage solution exploiting the filesystem and Elasticsearch for metadata in a raw format (e.g., original documents), 
and a relational DBMS for metadata in structured format, with respect to metadata type. This allows to take advantage of each storage mode's specific features. 

Metadata are also referred to through an \textit{mdRef} element in the manifest document. 
The \textit{XPTR} attribute of the \textit{mdRef} element is set with the metadata piece's URI, while its \textit{LABEL} attribute is set to the concatenation of the names of the two operations used for generating presentation metadata.

Table~\ref{tab:PresentationMetadata}  presents a short list of sample operations for generating presentation metadata. 
To the transformation operations, we add a special operation with a neutral effect named \textit{original version}. It is equivalent to a ``no transformation" operation on the original document.
We also define a presentation operation with a neutral effect, \textit{classic presentation}, which leaves the transformed document in its raw format. 

When these two special operations are applied to a document, presentation metadata are exactly identical to the original data.  
These two special, neutral operations actually allow to retain original documents in the data lake as presentation metadata.

\begin{table*}[hbt]
\caption{Transformation and presentation operations}
\label{tab:PresentationMetadata}
\begin{tabular} {|l|l|l|}
\hline
 \textbf{Transformation operation (to)} &  \textbf{Presentation operation (to)} & \textbf{Resulting metadata} \\
\hline
 Original version & Classic presentation & Original document \\
\hline
 Original version &  Term-frequency vector & Term-frequency vector  \\
\hline
 Original version & TF-IDF vector & TF-IDF vector \\
\hline
 Lemmatized version & Classic presentation & Lemmatized document \\
 \hline  
  Lemmatized version &  Term-frequency vector & Term-frequency vector   \\
  & & of lemmatized document \\
\hline
Lemmatized version & TF-IDF vector & TF-IDF vector of lemmatized document  \\
\hline
\end{tabular}
\end{table*}

Eventually, we assume that previsualization metadata represent a special case of presentation metadata. An example of presentation metadata can be obtained by filtering the original document on most frequent terms and presenting the result in a tag (term) cloud.

\subsubsection{Inter-document Metadata Extraction}

Some physical links can be automatically generated, e.g., the Apache Tika framework \cite{ApacheTika} permits to automatically detect document MIME type and language \cite{Quix2016}.  
Once the MIME type or language is detected, documents of same type or language can be considered physically linked, respectively.

However, other physical links must be defined through human intervention. It is the case of links expressing documents belonging to the same business category. For example, in a corpus composed of a company's annual reports, the department to which each document belongs must be defined either by the document's folder name or by a tag.

Finally, we propose to generate logical links by computing document similarity measures representing the similarity strength between each couple of documents. Some examples of textual data similarity measures include the cosine similarity \cite{Allan2000}, the chi-square similarity \cite{Ibrahimov2002} and Spearman's Rank Correlation Coefficient \cite{Kilgarriff2001}. Such similarity measures have a high score for documents sharing the same terms and a low score for documents using different terms.
Moreover, many different logical links can be obtained by applying the same measure on different presentations of the documents.

\subsubsection{Global Metadata}
Global metadata generation generally needs human intervention, because such metadata are domain-specific and must be designed by a domain expert  \cite{Quix2016}. 
Global metadata may also be derived from pre-existing metadata, e.g., a list of stopwords can easily be found on the Web and then complemented or reduced.

\section{\uppercase{Proof of Concept}}
\label{sec:POC}

Since there is no such systems as ours currently available, we put the metadata management system from Section~\ref{sec:MetadataManagement} in practice within a research project related to management sciences, and more specifically strategic marketing, as a proof of concept. This project is named COREL (``at the heart of customer relationship") and 
 aims to study, analyze and compare  customer policies between and within companies.   

We first present COREL's       
corpus in Section~\ref{sec:presCODAL}. 
Section~\ref{sec:architecture} is dedicated to the corresponding data lake's architecture.
Then, we explain in Section~\ref{sec:metadataExtraction} how we build the metadata system. Finally, Section~\ref{sec:queries} shows how our metadata system is exploited to perform analyses. 

\subsection{COREL Corpus}
\label{sec:presCODAL}

The COREL 
corpus is composed of 101 textual documents 
from 12 different companies from various business domains. Documents are categorized in 3 business categories: interviews, annual reports and press articles. The documents bear 2 different MIME types (application/pdf and application/vnd.openxmlformats-officedocument.wordprocessingml.document) and 2 different languages (French and English).

Although the COREL 
corpus does not fall in the big data category in terms of volume (the  size of  raw documents  is 0.15~GB), it does in terms of variety, with significant variations in annual reports with respect to companies, various sponsors in interviews (CEOs, marketers...) and a variety of press releases. 

\subsection{Data Lake Architecture}
\label{sec:architecture}
\begin{figure*}[hbt] 
 \centering
 \includegraphics[width=11cm]{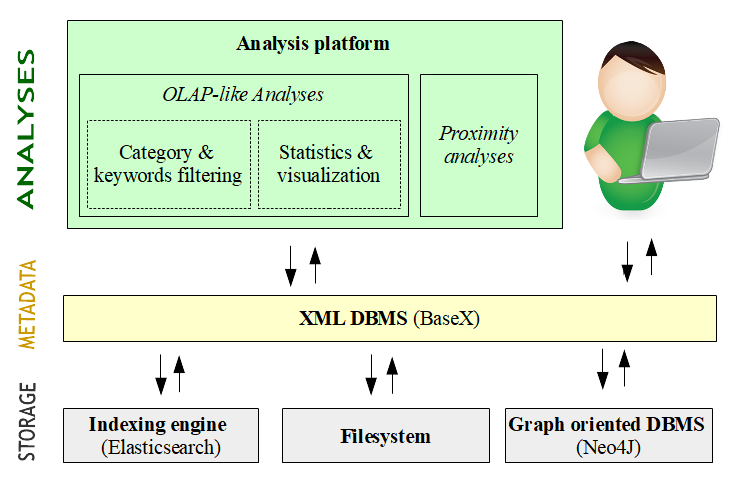} 
 \caption{Architecture of CODAL}
 \label{fig:architecture}
\end{figure*}

To allow various analyses on the COREL 
corpus, we build a data lake called CODAL (COrel DAta Lake) 
that is composed of one single data pond, since the whole corpus is purely textual. 
To fit CODAL within our metadata management system, we adopt the architecture and technologies shown in Figure~\ref{fig:architecture}. This architecture conforms to the storage-metadata-analysis system presented in Section~\ref{sec:mdsys}. 

The lowest-level component in this architecture is dedicated to document and metadata storage. It is hybrid and exploits the following technologies: the filesystem to store presentation metadata, Elasticsearch to index presentation metadata and Neo4j to store logical links.
To speed up queries, we plan to replace the filesystem by a relational DBMS, but we currently keep it as is for simplicity's sake.

The second component stores XML manifest documents that contain atomic metadata, physical links and pointers to non-atomic metadata. The set of XML manifest documents is stored in BaseX, an XML-native DBMS \cite{BaseX}.

Finally, the top-level component in the CODAL architecture is a layer that allows OLAP-like analyses through a Web platform.  In addition, CODAL users can access the metadata system through  BaseX to execute ad-hoc queries or to extract data.

\subsection{Metadata Extraction}
\label{sec:metadataExtraction}
The metadata extraction process is composed of two phases. The first consists in generating the set of XML manifest documents (Section~\ref{sec:intraExt}) and all relevant metadata, while the second is logical link generation (Section~\ref{sec:physExt}). 

\subsubsection{Intra-document Metadata and Physical Links}
\label{sec:intraExt}
For each document, 
we generate an XML manifest document bearing the format defined in Section~\ref{sec:manifest}. 
Properties are an identifier,  the document's title, its creator, the creation and the last modification date. These metadata are then stored as elements in the manifest first \textit{dmdSec} section/\textit{mdWrap} subsection.

Presentation metadata are each generated by applying a transformation followed by a presentation operation. The different transformation operations applied are stopwords removal, lemmatization, filtering on a dictionary and preservation of the original version. The operations retained for presentation are term-frequency vector format, TF-IDF vector format and classic presentation (raw format).

Once presentation metadata are generated, they are stored in the filesystem as either simple text files or in CSV format (for TF-IDF or term-frequency vectors).
They are then referred to in the corresponding XML manifest document. 
Presentation metadata in classic presentation are also indexed in Elasticsearch. 

Presentation metadata extraction is achieved using global metadata such as a list of stopwords and a dictionary. These global metadata are manually created and then referred to in the global XML manifest.

In the context of the COREL 
project, we retain four physical links: belonging to the same company, document type (business category), MIME type and language. The first two links are obtained via the documents' directory structure in the filesystem. The last two are extracted with Apache Tika.    

\begin{figure*}[hbt]
 \centering
 \includegraphics[width=12cm]{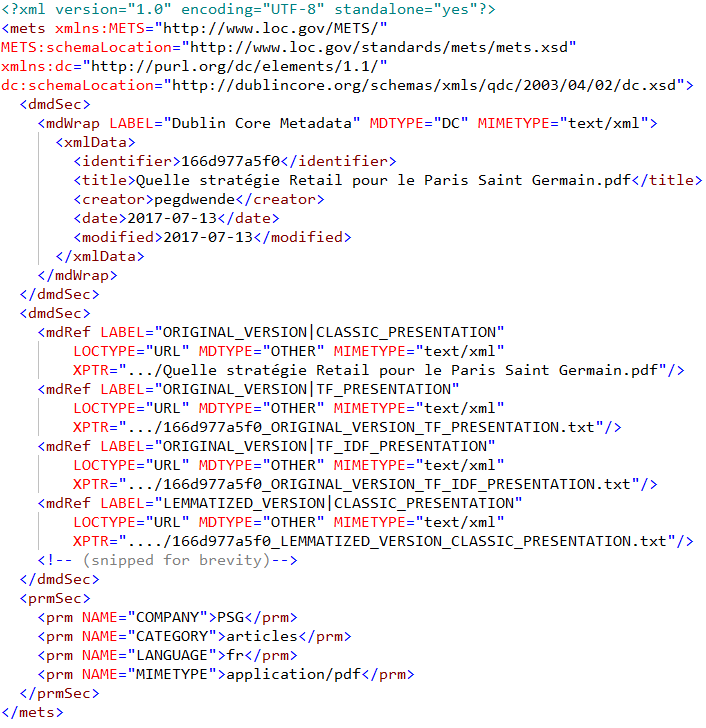} 
 \caption{Sample XML manifest}
 \label{fig:manifestExample}
\end{figure*}

Once all the XML manifest documents are created, they are inserted in a BaseX database. 
Figure~\ref{fig:manifestExample} shows an example of XML manifest document. 

\subsubsection{Logical links}
\label{sec:physExt}
To generate logical links, we calculate  the cosine similarity of each couple of documents using as presentation metadata TF-IDF or term-frequency vectors, since the cosine similarity can only be calculated with such vectors. 

To store the generated measures, a node is created in Neo4j for each document. The different measures are then integrated into the database as edges carrying the similarity's strength. Each edge is named by concatenating the presentation metadata piece's name with the similarity measure's name.

\subsection{Possible Queries and Analyses}
\label{sec:queries}
Authorized users can freely access the CODAL metadata management system to perform either ad-hoc queries or analyses through the BaseX and the Neo4j DBMSs. All the generated metadata can be filtered, aggregated and extracted through these interfaces. However, some advanced data management skills are required, which are not possessed by researchers in management sciences.

Thence, we developed an intuitive platform to allow some recurrent analyses, i.e., OLAP-like analyses (Section~\ref{sec:olapAnalyses}), document proximity analyses (Section~\ref{sec:linkAnalyses})  and document highlights (Section~\ref{sec:highlights}).

\subsubsection{OLAP-like Analyses}
\label{sec:olapAnalyses}
\begin{figure*}[hbt]
 \centering
 \includegraphics[width=16cm]{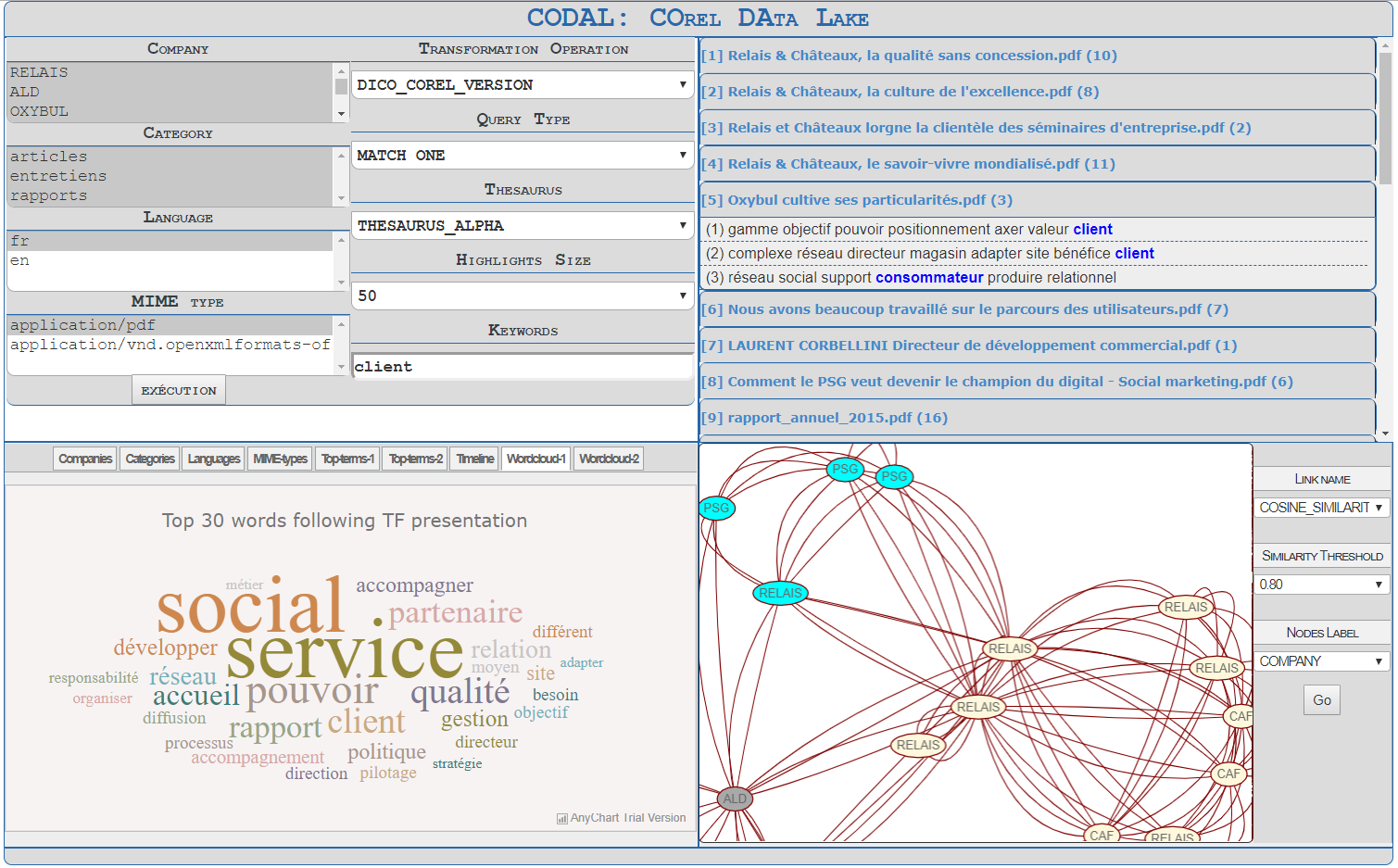} 
 \caption{CODAL analysis interface}
 \label{fig:interface}
\end{figure*}

These analyses are done through the left-hand side of our Web interface (Figure~\ref{fig:interface}). A set of multiple choice boxes allows documents filtering and aggregation with respect to physical links. 

Documents can also be filtered and aggregated using keyword-based queries. Here, a transformation operation must be specified and filtering operates using the classic document presentation  resulting from this transformation. Another option allows to extend keyword filtering with all synonyms of the given terms through a thesaurus. 
The filtering options serve as analysis axes that are much similar to OLAP dimensions. Thus, filtering the corpus on a set of physical links is similar to OLAP slice and dice operations.

The characteristics of the aggregated documents are provided through statistics and visualizations. These visualizations can be compared to OLAP measures, because they also provide some aggregated information about documents. We propose four types of visualizations:
    1) distribution of documents through clusters induced by physical links, e.g., displaying what companies provide documents in English (filter documents on the English language and then observe the results' distribution with respect to companies);
    2)~timeline created from the document's creation or last modification date, which provides a temporal distribution of documents;
    3) most common terms (plotted as a bar graph);
    4) average term frequencies (depicted as a tag cloud).

The left-bottom part of Figure~\ref{fig:interface} displays a visualization of the most common terms in the selected documents with a tag cloud where the size of terms represents their weight.

As the analysis platform is a Web application, different filtering can be performed simultaneously in different windows, to compare a given visualization with respect to various filter values. 
Similarly, the same filter can be set to observe simultaneously two or more visualizations. 

\subsubsection{Proximity Analyses}
\label{sec:linkAnalyses}
The right-bottom side of the analysis platform  (Figure~\ref{fig:interface}) is dedicated to  proximity analyses on, and  proximity visualizations of, selected documents.
To show what documents are similar or different, we cluster them automatically by applying the Walktrap community detection algorithm \cite{Pons2006} on the sub-graph induced by the selected documents, given a similarity measure and a user-defined threshold. Walktrap's time complexity is $O(n^{2}~log~n)$ in most cases, where $n$ is the number of nodes.
The result of clustering allows identifying the documents that form groups with strong internal links, while links with documents outside of the group are weak. 

Then, we represent the documents in a graph with a different color for each cluster. In Figure~\ref{fig:interface}, we observe in the yellow-colored cluster documents from two distinct companies. This can be interpreted as a similarity in the vocabulary used by these companies in some of their documents and, thence,  a potential similarity in their marketing strategy. 

Another relevant technique for proximity analysis is documents centrality calculation \cite{Farrugia2016}, which can be applied in the context of textual documents to identify the documents bearing a specific or common vocabulary. Documents with a specific vocabulary are weakly linked to others, which implies a low centrality. In contrast, documents with a common vocabulary are highly connected together and have a high centrality. 

Centrality can also be interpreted as the document's importance in the graph. Thus, documents with a high centrality can be considered essential because they are involved in a large number of links. Such documents should then be handled carefully, so as not to be destroyed. 
However, this technique is not implemented yet in our analysis platform.

\subsubsection{Highlights}
\label{sec:highlights}

These visualizations are provided in the right-top part of the CODAL analysis interface (Figure~\ref{fig:interface}). After keyword-based filtering, the corresponding documents are listed. 
In addition, we show, for each resulting document, a set of highlights where the keywords appear. This constitutes a summary of the document "around" one or more keywords. 

Moreover, advanced options in the left-top side of the interface allow to customize highlight display. The \textit{highlights size} option can be used to increase or decrease the highlight's length. The \textit{thesaurus} option allows to expand the given keywords with all their synonyms, with the help of a previously selected thesaurus.
For example, we observe in Figure~\ref{fig:interface} that a query on the term ``client" also returns a highlight on the term ``consommateur", i.e, consumer in French.

\section{\uppercase{Conclusion}}
\label{sec:Conclusion}

We propose in this article the first, to the best of our knowledge, complete methodological approach for building a metadata management system for data lakes or data ponds storing textual documents.
To avoid the data swamp syndrome, we  
identify relevant metadata extraction, structuring, storage  and processing techniques and tools. 
We notably distinguish three types of metadata, each of which having its own extraction and storage techniques: intra-document metadata, inter-document metadata and global, semantic metadata (which we introduce). Eventually, we extend the XML manifest metadata representation to suit textual document-related metadata storage.

We apply and validate the feasibility of our metadata management system on a real-life textual corpus to build the CODAL data lake.  As a result, non-specialist users (i.e., with no data management knowledge) can perform OLAP-like analyses. Such analyses consist in filtering and aggregating the corpus with respect one or more terms, and in navigating through visualizations that summarize the filtered corpus. Users can also cluster documents to identify  groups of similar documents. 

In future works, we first plan to replace the filesystem by a relational DBMS to store structured presentation metadata, and thus allow easier and faster queries and analyses.
We shall also improve our platform by integrating centrality analyses (Section~\ref{sec:linkAnalyses}). 
Finally, since our current test corpus is small-sized, we plan to apply our method on a bigger one from a new project in management sciences and test its scalability.

Moreover, our objective is to turn the specific CODAL platform into a generic (i.e., not tied to the COREL project) analysis platform that implements the metadata management system we propose, and make it available to the community.  This would allow non-computer scientists to easily exploit any textual data pond. 

Eventually, in the long run, we aim at designing a metadata management system that would help querying data ponds storing different types of data (structured, semi-structured, unstructured -- textual and possibly multimedia) altogether.

\section*{\uppercase{Acknowledgements}}
The research accounted for in this paper has been funded by the Universit\'e Lumi\`ere Lyon~2 through the COREL project. The authors would like to thank Angsoumailyne Te and Isabelle Prim-Allaz, from the COACTIS management science research center, for their constant input and feedback.

\bibliographystyle{apalike}
{\small
\bibliography{biblio}}

\begin{thebibliography}{}

\bibitem[Allan et~al., 2000]{Allan2000}
Allan, J., Lavrenko, V., Malin, D., and Swan, R. (2000).
\newblock Detections, bounds, and timelines: Umass and tdt-3.
\newblock In {\em Topic Detection and Tracking Workshop (TDT-3), Vienna, VA,
  USA}, pages 167--174.

\bibitem[Ansari et~al., 2018]{Ansari2018}
Ansari, J.~W., Karim, N., Decker, S., Cochez, M., and Beyan, O. (2018).
\newblock {Extending Data Lake Metadata Management by Semantic Profiling}.
\newblock In {\em 2018 Extended Semantic Web Conference (ESWC 2018), Heraklion,
  Crete, Greece}, ESWC, pages 1--15.

\bibitem[{BaseX~GmbH}, 2018]{BaseX}
{BaseX~GmbH} (2018).
\newblock {BaseX -- The XML Framework}.
\newblock http://basex.org/.

\bibitem[Beheshti et~al., 2017]{Beheshti2017}
Beheshti, A., Benatallah, B., Nouri, R., Chhieng, V.~M., Xiong, H., and Zhao,
  X. (2017).
\newblock {CoreDB: a Data Lake Service}.
\newblock In {\em 2017 ACM on Conference on Information and Knowledge
  Management (CIKM 2017), Singapore, Singapore}, ACM, pages 2451--2454.

\bibitem[Dixon, 2010]{Dixon2010}
Dixon, J. (2010).
\newblock {Pentaho, Hadoop, and Data Lakes}.
\newblock
  https://jamesdixon.wordpress.com/2010/10/14/pentaho-hadoop-and-data-lakes/.

\bibitem[{Dublin~Core~Metadata~Initiative}, 2018]{DublinCore}
{Dublin~Core~Metadata~Initiative} (2018).
\newblock {Dublin Core}.
\newblock http://dublincore.org/.

\bibitem[Elastic, 2018]{Elasticsearch}
Elastic (2018).
\newblock Elasticsearch.
\newblock https://www.elastic.co.

\bibitem[Fang, 2015]{Fang2015}
Fang, H. (2015).
\newblock {Managing Data Lakes in Big Data Era: What's a data lake and why has
  it became popular in data management ecosystem}.
\newblock In {\em 5th Annual IEEE International Conference on Cyber Technology
  in Automation, Control and Intelligent Systems (CYBER 2015), Shenyang,
  China}, IEEE, pages 820--824.

\bibitem[Farid et~al., 2016]{Farid2016}
Farid, M., Roatis, A., Ilyas, I.~F., Hoffmann, H.-F., and Chu, X. (2016).
\newblock {CLAMS: Bringing Quality to Data Lakes}.
\newblock In {\em 2016 International Conference on Management of Data (SIGMOD
  2016), San Francisco, CA, USA}, ACM, pages 2089--2092.

\bibitem[Farrugia et~al., 2016]{Farrugia2016}
Farrugia, A., Claxton, R., and Thompson, S. (2016).
\newblock {Towards Social Network Analytics for Understanding and Managing
  Enterprise Data Lakes}.
\newblock In {\em Advances in Social Networks Analysis and Mining (ASONAM
  2016), San Francisco, CA, USA}, IEEE, pages 1213--1220.

\bibitem[Fauduet and Peyrard, 2010]{Fauduet2010}
Fauduet, L. and Peyrard, S. (2010).
\newblock A data-first preservation strategy: Data management in spar.
\newblock In {\em 7th International Conference on Preservation of Digital
  Objects (SPAR 2010), Vienna, Autria}, pages 1--8.

\bibitem[Hai et~al., 2017]{Hai2016}
Hai, R., Geisler, S., and Quix, C. (2017).
\newblock {Constance: An Intelligent Data Lake System}.
\newblock In {\em 2016 International Conference on Management of Data (SIGMOD
  2016) ,San Francisco, CA, USA}, ACM Digital Library, pages 2097--2100.

\bibitem[Halevy et~al., 2016]{Halevy2016}
Halevy, A., Korn, F., Noy, N.~F., Olston, C., Polyzotis, N., Roy, S., and
  Whang, S.~E. (2016).
\newblock {Managing Google's data lake: an overview of the GOODS system}.
\newblock In {\em 2016 International Conference on Management of Data (SIGMOD
  2016), San Francisco, CA, USA}, ACM, pages 795--806.

\bibitem[Hultgren, 2016]{Hultgren2016}
Hultgren, H. (2016).
\newblock {\em {Data Vault modeling guide: Introductory Guide to Data Vault
  Modeling}}.
\newblock Genessee Academy, USA.

\bibitem[Ibrahimov et~al., 2002]{Ibrahimov2002}
Ibrahimov, O., Sethi, I., and Dimitrova, N. (2002).
\newblock {The Performance Analysis of a Chi-square Similarity Measure for
  Topic Related Clustering of Noisy Transcripts}.
\newblock In {\em 16th International Conference on Pattern Recognition, Quebec
  City, Quebec, Canada}, pages 285--288.

\bibitem[Inmon, 2016]{Inmon2016}
Inmon, B. (2016).
\newblock {\em {Data Lake Architecture: Designing the Data Lake and avoiding
  the garbage dump}}.
\newblock Technics Publications.

\bibitem[Kilgarriff, 2001]{Kilgarriff2001}
Kilgarriff, A. (2001).
\newblock {Comparing Corpora}.
\newblock {\em International Journal of Corpus Linguistics}, 6(1):97--133.

\bibitem[Klettke et~al., 2017]{Klettke2017}
Klettke, M., Awolin, H., St\"url, U., M\"uller, D., and Scherzinger, S. (2017).
\newblock {Uncovering the Evolution History of Data Lakes}.
\newblock In {\em 2017 IEEE International Conference on Big Data (BIGDATA
  2017), Boston, MA, USA}, pages 2462--2471.

\bibitem[Laskowski, 2016]{Laskowski2016}
Laskowski, N. (2016).
\newblock {Data lake governance: A big data do or die}.
\newblock
  https://searchcio.techtarget.com/feature/Data-lake-governance-A-big-data-do-or-die.

\bibitem[Linstedt, 2011]{Linstedt2011}
Linstedt, D. (2011).
\newblock {\em {Super Charge your Data Warehouse: Invaluable Data Modeling
  Rules to Implement Your Data Vault}}.
\newblock CreateSpace Independent Publishing.

\bibitem[Maccioni and Torlone, 2017]{Maccioni2017}
Maccioni, A. and Torlone, R. (2017).
\newblock {Crossing the finish line faster when paddling the data lake with
  KAYAK}.
\newblock {\em VLDB Endowment}, 10(12):1853--1856.

\bibitem[Madera and Laurent, 2016]{Madera2016}
Madera, C. and Laurent, A. (2016).
\newblock The next information architecture evolution: the data lake wave.
\newblock In {\em 8th International Conference on Management of Digital
  EcoSystems ({MEDES} 2016), Biarritz, France}, pages 174--180.

\bibitem[Miloslavskaya and Tolstoy, 2016]{Miloslavskaya2016}
Miloslavskaya, N. and Tolstoy, A. (2016).
\newblock {Big Data, Fast Data and Data Lake Concepts}.
\newblock In {\em 7th Annual International Conference on Biologically Inspired
  Cognitive Architectures (BICA 2016), NY, USA}, volume~88 of {\em Procedia
  Computer Science}, pages 1--6.

\bibitem[{Neo4J~Inc.}, 2018]{Neo4j}
{Neo4J~Inc.} (2018).
\newblock {The Neo4j Graph Platform}.
\newblock https://neo4j.com.

\bibitem[Nogueira et~al., 2018]{Nogueira2018}
Nogueira, I., Romdhane, M., and Darmont, J. (2018).
\newblock {Modeling Data Lake Metadata with a Data Vault}.
\newblock In {\em 22nd International Database Engineering and Applications
  Symposium (IDEAS 2018), Villa San Giovanni, Italia}, pages 253--261, New
  York. ACM.

\bibitem[Pons and Latapy, 2006]{Pons2006}
Pons, P. and Latapy, M. (2006).
\newblock {Computing Communities in Large Networks Using Random Walks}.
\newblock {\em Journal of Graph Algorithms and Applications}, 10(2):191--218.

\bibitem[Quix et~al., 2016]{Quix2016}
Quix, C., Hai, R., and Vatov, I. (2016).
\newblock {Metadata Extraction and Management in Data Lakes With GEMMS}.
\newblock {\em Complex Systems Informatics and Modeling Quarterly},
  (9):289--293.

\bibitem[Stein and Morrison, 2014]{Stein2014}
Stein, B. and Morrison, A. (2014).
\newblock {The enterprise data lake: Better integration and deeper analytics}.
\newblock {\em PWC Technology Forecast}, (1):1--9.

\bibitem[Suriarachchi and Plale, 2016]{Suriarachchi2016}
Suriarachchi, I. and Plale, B. (2016).
\newblock {Crossing Analytics Systems: A Case for Integrated Provenance in Data
  Lakes}.
\newblock In {\em 12th {IEEE} International Conference on e-Science (e-Science
  2016), Baltimore, MD, USA, October 23-27, 2016}, pages 349--354.

\bibitem[Terrizzano et~al., 2015]{Terrizzano2015}
Terrizzano, I., Schwarz, P., Roth, M., and Colino, J.~E. (2015).
\newblock {Data Wrangling: The Challenging Journey from the Wild to the Lake}.
\newblock In {\em 7th Biennial Conference on Innovative Data Systems Research
  (CIDR 2015), Asilomar, CA, USA}, pages 1--9.

\bibitem[{The~Apache~Software~Foundation}, 2018]{ApacheTika}
{The~Apache~Software~Foundation} (2018).
\newblock {Apache Tika -- a content analysis toolkit}.
\newblock https://tika.apache.org/.

\bibitem[{The~Library~of~Congress}, 2017]{TLoC2017}
{The~Library~of~Congress} (2017).
\newblock {METS: An Overview and Tutorial}.
\newblock http://www.loc.gov/standards/mets/METSOverview.v2.html.

\end{thebibliography}

\vfill
\end{document}